\documentclass[aps,prl,twocolumn,showpacs,preprintnumbers,amsmath,amssymb,floatfix,nofootinbib]{revtex4-1}
\usepackage{graphicx}

\usepackage{color}
\usepackage{colordvi}

\usepackage[normalem]{ulem}

\usepackage{color}

\newcommand{\beqn}{\begin{eqnarray}}
\newcommand{\eeqn}{\end{eqnarray}}
\newcommand{\be}{\begin{equation}}
\newcommand{\ee}{\end{equation}}
\newcommand{\eqn}[1]{(\ref{#1})}
\newcommand{\bel}[1]{\be\label{#1}}
\newcommand{\ba}{\begin{array}{c}}
\newcommand{\bat}{\begin{array}{cc}}
\newcommand{\ea}{\end{array}}

\newcommand{\bi}{\begin{itemize}}
\newcommand{\ei}{\end{itemize}}

\newcommand{\ket}{\,\rangle}
\newcommand{\bra}{\langle \,}

\newcommand{\Frac}[2]{\frac{\displaystyle #1}{\displaystyle #2}}
\newcommand{\no}{\nonumber}

\newcommand{\cO}{{\cal O}}

\newcommand{\mF}{\mathcal{F}}

\newcommand{\mL}{\mathcal{L}}

\newcommand{\mO}{\mathcal{O}}
\newcommand{\mP}{\mathcal{P}}

\newcommand{\lsim}{\stackrel{<}{_\sim}}

\begin{document}

\preprint{IFIC/15-74}
\preprint{FTUV-15-1012}
\preprint{IFT-UAM/CSIC-15-107}
\preprint{FTUAM-15-31}


\title{Low-energy signals of strongly-coupled electroweak symmetry-breaking scenarios}

\author{Antonio Pich${}^{1}$}
\author{Ignasi Rosell${}^{2}$}

\author{Joaqu\'{\i}n Santos${}^{1}$}

\author{Juan Jos\'e Sanz-Cillero${}^{3}$}

\affiliation{${}^1$ Departament de F\'\i sica Te\`orica, IFIC, Universitat de Val\`encia -- CSIC, Apt. Correus 22085, 46071 Val\`encia, Spain }

\affiliation{${}^2$  Departamento de Ciencias F\'\i sicas, Matem\'aticas y de la Computaci\'on,
Universidad CEU Cardenal Herrera, 46115 Alfara del Patriarca, Val\`encia, Spain}

\affiliation{${}^3$  Departamento de F\'{\i}sica Te\'orica and Instituto de F\'{\i}sica Te\'orica, IFT-UAM/CSIC, Universidad Aut\'onoma de Madrid, Cantoblanco, 28049 Madrid, Spain}

\begin{abstract}
The non-observation of new particles at the LHC suggests the existence of a mass gap above the electroweak scale. This situation is adequately described through a general 
electroweak effective theory with the established fields and Standard Model symmetries.
Its couplings contain all information about the unknown short-distance dynamics which is accessible at low energies. We consider a generic strongly-coupled scenario of electroweak symmetry breaking, with heavy states above the gap, and analyze the imprints that its lightest bosonic excitations leave on the effective Lagrangian couplings. Different quantum numbers of the heavy states imply different patterns of low-energy couplings, with characteristic correlations which could be identified in future data samples.
The predictions can be sharpened with mild assumptions about the
ultraviolet behaviour of the underlying fundamental theory.
\end{abstract}



\maketitle

\vspace{-5cm}

\section{Introduction}

The LHC data have confirmed the validity of the Standard Model (SM) framework at the electroweak scale, including the existence of a light Higgs boson \cite{Pich:2015tqa}. Moreover, the absence of signals of new phenomena pushes at higher energies the scale of hypothetical new dynamics, beyond the TeV in most scenarios. While there are many profound reasons to think that the SM is not the ultimate fundamental theory and new-physics should indeed exist,  a mass gap between the electroweak and new-physics scales appears to be present. Below the gap, the only possible signals of the high-energy dynamics are hidden in the couplings of the low-energy electroweak effective theory (EWET), which can be
tested through scattering amplitudes involving only SM particles.

The EWET contains the most general Lagrangian, built with the established fields, compatible with the SM gauge symmetries and the known pattern of electroweak symmetry breaking (EWSB):
$G\equiv SU(2)_L\otimes SU(2)_R\rightarrow H\equiv SU(2)_{L+R}$, giving rise to three Goldstone bosons which account for the longitudinal polarizations of the $W^\pm$ and $Z$ gauge bosons.
Although most recent works concentrate in the particular case of a linear realization of EWSB, assuming the Higgs boson to be part of a $SU(2)_L$ doublet, as in the SM,
we will consider the more general non-linear framework with a singlet Higgs field $h(x)$, unrelated to the Goldstone triplet $\vec\varphi(x)$. The Goldstone fields are parametrized through the canonical $G/H$ coset representative \cite{Coleman:1969sm,Callan:1969sn,Ecker:1988te,Ecker:1989yg} $u(\varphi) =\exp{(\frac{i}{2}\,\vec\sigma\vec\varphi/v)}$, with $v= 246~\mathrm{GeV}$ the electroweak scale.
Under chiral transformations $g\equiv (g_L^{\phantom{\dagger}},g_R^{\phantom{\dagger}})\in G$,
$u(\varphi)\to g_L^{\phantom{\dagger}} u(\varphi) g_h^\dagger(\varphi,g) = g_h^{\phantom{\dagger}}(\varphi,g)
u(\varphi) g^\dagger_R$, with $g_h^{\phantom{\dagger}}(\varphi,g)\equiv g_h^{\phantom{\dagger}}\in H$ \cite{Pich:2012dv,Pich:2013fea,Pich:2012jv}.

The effective Lagrangian is organized as a low-energy expansion in powers of momenta (and symmetry breakings): $\mL =\sum_d \mL_d$, with $\mL_d$ of $\cO(p^d)$.
At lowest-order (LO), $\cO(p^2)$, it contains the renormalizable massless (unbroken) SM Lagrangian plus the Goldstone term
\bel{eq:L_phi}
\mL_2(\varphi)\, =\, \frac{v^2}{4}\,\langle D_\mu U^\dagger D^\mu U\rangle\, =\,
\frac{v^2}{4}\,\langle u_\mu u^\mu\rangle\, ,
\ee
where $U=u^2\to g_L^{\phantom{\dagger}} U g^\dagger_R$, $u_\mu = i\, u\, (D_\mu U)^\dagger u = u_\mu^\dagger\to g_h^{\phantom{\dagger}} u_\mu g_h^{\dagger}$ and $\langle O \rangle$ denotes the 2-dimensional trace of $O$. Eq.~\eqn{eq:L_phi} is the universal Goldstone Lagrangian associated with the symmetry breaking $G\to H$. The same structure with $v\to f_\pi$ and $\vec\varphi\to\vec\pi$ governs the low-energy pion dynamics in 2-flavour
Quantum Chromodynamics (QCD) \cite{Pich:1998xt}.
The covariant derivative $D_\mu U = \partial_\mu U - i \hat W_\mu U + i U \hat B_\mu$ couples the Goldstones to external $SU(2)_{L,R}$ gauge sources, making the Lagrangian formally invariant under local $G$ transformations. The identification with the SM gauge fields, $\hat W_\mu = -\frac{g}{2}\,\vec\sigma\vec W_\mu$ and $\hat B_\mu = -\frac{g'}{2}\,\sigma_3 B_\mu$, breaks explicitly the $SU(2)_R$ symmetry while preserving the $SU(2)_L\otimes U(1)_Y$ SM symmetry.

The non-linear Lagrangian~\eqn{eq:L_phi} contains arbitrary powers of Goldstone fields, compensated by corresponding powers of the scale $v$. Since $h$ and $\vec\varphi$ are assumed to have similar origins, powers of $h/v$ do not increase either the chiral dimension. Therefore, in the EWET the term~\eqn{eq:L_phi} must be multiplied by an arbitrary function of the Higgs field $\mF_u(h/v)$~\cite{Grinstein:2007iv}.
$\mL_2$ includes in addition the kinetic Higgs Lagrangian, with mass $m_h$, and a scalar potential $V(h/v)$ containing arbitrary powers of $h/v$.

At the next-to-leading order (NLO), one must consider one-loop contributions \cite{Guo:2015isa} from the LO Lagrangian plus $\cO(p^4)$ local structures. Restricting the analysis to bosonic fields and assuming that parity ($\mP:\; L \leftrightarrow R$) is a good symmetry of the EWSB sector,
\bel{eq:NLO}
\mL_4^{\mathrm{Bosonic}}\, =\, \sum_i \mF_i(h/v)\,\mO_i\, ,
\ee
with $\mF_i(h/v)$ arbitrary polynomials of $h/v$.
A basis of NLO operators $\mO_i$ was written down by Longhitano in the Higgsless theory~\cite{Longhitano:1980iz,Longhitano:1980tm,Appelquist:1980vg},
which must now be expanded with structures involving explicitly the Higgs field~\cite{Buchalla:2012qq,Alonso:2012px,Buchalla:2013rka}.
Table~\ref{tab:Oi}  collects the $\mP$-even operators which preserve custodial symmetry ($H$).
For convenience, the left and right field-strength tensors have been re-written in terms of $f_\pm^{\mu\nu}\equiv u^\dagger \hat W^{\mu\nu} u \pm u\, \hat B^{\mu\nu} u^\dagger$,
which transform as triplets under $G$:
$f_\pm^{\mu\nu} \to g_h^{\phantom{\dagger}} f_\pm^{\mu\nu} g_h^{\dagger}$.

All information on the underlying short-distance dynamics is encoded in the low-energy couplings (LECs) multiplying these operators. They can be accessed experimentally through precision measurements of anomalous triple and quartic gauge couplings, scattering amplitudes of longitudinal gauge bosons, Higgs couplings, etc. Once a clear pattern of LECs would emerge from the data, one would like to identify the physics originating these effects at high scales. In this letter, we analyze the low-energy signals of generic massive bosonic states, following the successful methodology developed in QCD to determine the LECs of chiral perturbation theory \cite{Ecker:1988te,Ecker:1989yg,Pich:2002xy,Cirigliano:2006hb}.


\begin{table}[tb]
{\renewcommand{\arraystretch}{1.2}
\begin{tabular}{|c|c|}
\hline
$\mO_1 =\frac{1}{4}\,\bra f_+^{\mu\nu} f^+_{\mu\nu}
- f_-^{\mu\nu} f^-_{\mu\nu}\ket$ &
$\mO_6 =\frac{1}{v^2}\, (\partial_\mu h)(\partial^\mu h)\,\bra u_\nu u^\nu \ket$
\\
$\mO_2 =\frac{1}{2}\,\bra f_+^{\mu\nu} f^+_{\mu\nu}
+ f_-^{\mu\nu} f^-_{\mu\nu}\ket$ &
$\mO_7 =\frac{1}{v^2}\, (\partial_\mu h)(\partial_\nu h) \,\bra u^\mu u^\nu \ket$
\\
$\mO_3 =\frac{i}{2}\,  \bra {f}_+^{\mu\nu} [u_\mu, u_\nu] \ket$ &
$\mO_8 =\frac{1}{v^4}\, (\partial_\mu h)(\partial^\mu h)(\partial_\nu h)(\partial^\nu h)$
\\
$\mO_4 =\bra u_\mu u_\nu\ket \, \bra u^\mu u^\nu\ket $ &
$\mO_9 =\frac{1}{v}\, (\partial_\mu h) \,\bra f_-^{\mu\nu}u_\nu \ket$
\\
$\mO_5 =\bra u_\mu u^\mu\ket^2$ &
\\
\hline
\end{tabular}}
\caption{$\mO(p^4)$ $\mP$-even bosonic operators of the EWET.}
\label{tab:Oi}
\end{table}

\section{Resonance effective theory}

Above the energy gap, the underlying dynamical theory (either effective or fundamental) contains the SM fields plus heavier degrees of freedom. We will concentrate here in new massive bosonic states (resonances R), postponing to future work the more involved study of additional fermions \cite{BigPaper}. Technicolor~\cite{Weinberg:1979bn,Weinberg:1975gm,Susskind:1978ms} and Walking Technicolor~\cite{Holdom:1984sk,Appelquist:1986an,Yamawaki:1985zg}, the most studied strongly-coupled models, predict the existence of bound states in the TeV range. The analysis of the oblique $S$ and $T$ parameters in near-conformal theories requires masses of at least a few TeV for the lightest spin-1 resonances~\cite{Pich:2012dv,Pich:2013fea,Orgogozo:2011kq,Foadi:2007ue,Foadi:2007se}. Likewise, composite fermions (technibaryons) are expected to appear above the TeV scale~\cite{Ryttov:2008xe}.

The exchange of heavy fields (propagators) contributes to Green functions with only light fields in the external lines. At low energies, these contributions are suppressed by powers of momenta over the heavy masses and, therefore, those states closer to the gap (the lightest heavy states) will dominate.

Let us consider an effective Lagrangian containing the SM fields coupled to the lightest scalar, pseudoscalar, vector and axial-vector colour-singlet resonance multiplets $S$, $P$, $V^{\mu\nu}$ and $A^{\mu\nu}$, transforming as $SU(2)_{L+R}$ triplets, {\it i.e.},
$R\to g_h^{\phantom{\dagger}} R\, g_h^{\dagger}$, and the corresponding singlet states $S_1$, $P_1$, $V^{\mu\nu}_1$ and $A^{\mu\nu}_1$ ($R_1\to R_1$). Since we are interested in the low-energy implications, we only need to 
keep those structures with the lowest number of resonances and derivatives.
At LO, the most general $\mP$-even bosonic interaction with at most one resonance field,
invariant under the symmetry group $G$, has the form:
\begin{eqnarray}\label{eq:Lagrangian}
\mathcal{L}\; &=\; &
\frac{v^2}{4}\,\bra \! u_\mu u^\mu \!\ket \left( 1 + \frac{2\,\kappa_W^{\phantom{2}}}{v}\, h
+ \frac{4 c_d}{\sqrt{2}v^2}\, S_1\right)
+ \lambda_{hS_1} v\, h^2 S_1
\nonumber \\ &&\hskip -.8cm\mbox{} +
\frac{F_V}{2\sqrt{2}}\, \bra \!V_{\mu\nu} f^{\mu\nu}_+ \!\ket
+ \frac{i\, G_V}{2\sqrt{2}}\, \bra \! V_{\mu\nu} [u^\mu, u^\nu] \!\ket
 + \frac{F_A}{2\sqrt{2}}\, \bra \!A_{\mu\nu} f^{\mu\nu}_- \!\ket
\nonumber \\ &&\hskip -.8cm\mbox{} +
\sqrt{2}\, \lambda_1^{hA}\,  \partial_\mu h \, \bra \! A^{\mu \nu} u_\nu \!\ket
+ \frac{d_P}{v}\, (\partial_\mu h)\,\bra P u^\mu\ket
\, .
\end{eqnarray}
In the first term we have included the linear Higgs coupling to the Goldstones which for $\kappa_W=1$ reproduces the gauge coupling of the SM Higgs. We have only made other Higgs couplings explicit when they give rise to new operators (all couplings must be actually understood as functions of $h/v$). The number of chiral structures has been reduced through field redefinitions, partial integration, equations of motion and algebraic identities \cite{BigPaper}. Notice that the singlet vector, axial-vector and pseudoscalar fields, and the scalar triplet cannot couple to the Goldstones and gauge bosons at this chiral order.
The spin-1 fields are described with antisymmetric tensors because this leads to a simpler formalism, avoiding mixings with the Goldstones, and better ultraviolet properties \cite{Ecker:1988te,Ecker:1989yg}.

The low-energy EWET of the Lagrangian~\eqn{eq:Lagrangian} is formally obtained, integrating out the heavy fields from the generating functional and expanding the resulting non-local action in powers of momenta over the heavy scales. At LO (tree-level exchanges) this is easily achieved, using the (power-expanded) equations of motion to express the heavy fields in terms of the SM ones, and substituting back those expressions in the resonance Lagrangian \cite{Ecker:1988te}.
Writing the linear resonance couplings in the form
\bel{eq:LR}
\mL_R\, =\, \langle V_{\mu\nu} \,\chi_V^{\mu\nu}\rangle + \langle A_{\mu\nu} \, \chi_A^{\mu\nu}\rangle + \langle P\,\chi_P^{\phantom{\mu}}\rangle + S_1 \, \chi_{S_1}^{\phantom{\mu}}\, ,
\ee
the result can be expressed in terms of the chiral structures $\chi_R^{\mu\nu}, \chi_R^{\phantom{\mu}}$, which only involve SM fields and can be directly read from Eq.~\eqn{eq:Lagrangian} \cite{BigPaper}:
\beqn\label{eq:DL2}
\Delta \mL_{R}^{\cO(p^4)} & =&
\sum_{R=V,A} \frac{1}{M_R^2} \, \left(
\frac{1}{2}\,\bra \chi_R^{\mu\nu}\ket \bra \chi_{R\,\mu\nu}^{\phantom{\mu}}\ket
- \bra  \chi_R^{\mu\nu}\, \chi_{R\,\mu\nu}^{\phantom{\mu}}\ket \right)
\no\\ &+&
\frac{1}{2M_P^2} \,
 \left(  \bra  \chi_P^{\phantom{\mu}}\, \chi_P^{\phantom{\mu}}\ket   -\frac{1}{2}\bra \chi_P^{\phantom{\mu}}\ket^2\right)
\, +\, \Frac{ \chi_{S_1}^2}{2M_{S_1}^2}
\, .
\eeqn
Decomposing this expression in terms of the operator basis $\mO_i$, one obtains the predictions for the EWET LECs given in Table~\ref{tab:LECs}. The $S_1$-exchange term generates in addition an $\mO(p^4)$ correction to the lowest-order EWET Lagrangian:
\bel{eq:S1-exchange}
\Delta\mL_{S_1}^{(2)}\, =\,\frac{\lambda_{hS_1} v}{2 M_{S_1}^2}\; h^2 \left\{ \lambda_{hS_1} v\, h^2 + \sqrt{2}  c_d\,  \bra u_\mu u^\mu\ket\right\}\, .
\ee
%

\begin{table}[tb]
{\renewcommand{\arraystretch}{1.6}
\begin{tabular}{|cccc|}
\hline
$\mF_1\; =$ & $\frac{F_A^2}{4M_A^2}- \frac{F_V^2}{4M_V^2}$ & = &
$-\frac{v^2}{4}\,\left(\frac{1}{M_V^2}+\frac{1}{M_A^2}\right)$
\\
$\mF_2\; =$ & $-\frac{F_A^2}{8M_A^2}- \frac{F_V^2}{8M_V^2}$ &=&
$-\frac{v^2 (M_V^4+M_A^4)}{8 M_V^2M_A^2 (M_A^2-M_V^2)}$
\\
$\mF_3\; =$ & $-  \frac{F_VG_V}{2M_V^2}$ &=& $-\frac{v^2}{2M_V^2}$
\\
$\mF_4\; =$ & $\frac{G_V^2}{4 M_V^2}$ &=&
$\frac{(M_A^2-M_V^2) v^2}{4 M_V^2 M_A^2}$
\\
$\mF_5\; =$ & $\frac{c_{d}^2}{4M_{S_1}^2} -\frac{G_V^2}{4M_V^2}$ &=&
$\frac{c_{d}^2}{4M_{S_1}^2} -\frac{(M_A^2-M_V^2) v^2}{4 M_V^2 M_A^2}$
\\
$\mF_6\; =$ & $-\frac{(\lambda_1^{hA})^2v^2}{M_A^2}$ &=&
$-\frac{M_V^2 (M_A^2-M_V^2) v^2}{M_A^6}$
\\
$\mF_7\; =$ & $\frac{d_P^2}{2 M_P^2}+ \frac{(\lambda_1^{hA})^2v^2}{M_A^2}$ &=&
$\frac{d_P^2}{2 M_P^2}+\frac{M_V^2 (M_A^2-M_V^2) v^2}{M_A^6}$
\\
$\mF_8\; =$ & $0$ && 
\\
$\mF_9\; =$ & $- \frac{F_A \lambda_1^{hA} v}{M_A^2}$ &=& $-\frac{M_V^2 v^2}{M_A^4}$
\\ \hline
\end{tabular}
}
\caption{Predicted LECs from resonance exchange. The r.h.s. expressions include short-distance constraints.}
\label{tab:LECs}
\end{table}

The predicted spin-1 contributions are independent of the (antisymmetric) formalism adopted to describe the fields. The same results are obtained using Proca fields or a hidden-gauge formalism~\cite{Bando:1987br,Casalbuoni:1988xm}, once a proper ultraviolet (UV) behaviour is required (physical Green functions should not grow at large momenta) \cite{Ecker:1989yg,BigPaper}.

The LECs $\mF_i(h/v)$ are related to the Higgsless Longhitano's Lagrangian
couplings~\cite{Longhitano:1980iz,Longhitano:1980tm,Herrero:1993nc}, frequently used in previous literature,
through $a_1=\mF_1(0)$, $a_2-a_3=\mF_3(0)$, $a_4=\mF_4(0)$ and $a_5=\mF_5(0)$;
all other combinations being zero within our approximations.
Possible low-energy contributions from exotic $J^{PC}=1^{+-}$ heavy states haven been analyzed in Ref.~\cite{Cata:2014fna}.

\section{Short-distance constraints}

The integration of the heavy fields has generated a definite pattern of LECs in terms of the resonance masses and couplings.
Even if these parameters are unknown, the particular quantum numbers of a given intermediate resonance give rise to clear correlations among several LECs, which could be phenomenologically identified in future data samples.
These predictions can be sharpened, assuming a given UV behaviour of the underlying fundamental theory. Imposing the expected fall-off at large momenta of specific Green functions, one obtains constraints on the resonance parameters which are valid in broad classes of dynamical theories.

Taking functional derivatives of the action with respect to the external sources $\hat W^\mu, \hat B^\mu$, one defines the corresponding left/right (vector/axial) currents. The 2-Goldstone matrix element of the vector current is characterized by the vector form factor
\bel{eq:VFF}
F^V_{\varphi\varphi}(s)\, =\, 1 + \frac{F_V G_V}{v^2}\,\frac{s}{M_V^2-s}\, .
\ee
Requiring $F^V_{\varphi\varphi}(s)$ to vanish at infinite momentum transfer gives the relation \cite{Ecker:1988te,Ecker:1989yg}
\bel{eq:FVGV}
F_V G_V\, =\, v^2\, .
\ee
Applying a similar reasoning to the Higgs-Goldstone matrix element of the axial current, one obtains \cite{Pich:2012dv,Pich:2013fea}:
\bel{eq:FAkappa}
F_A \lambda_1^{hA}\, =\, \kappa_W^{\phantom{2}} v\, .
\ee

The two-point correlator of a left and a right currents,
$\Pi_{LR}(s)\! =\! \Pi_{VV}(s)\! -\!\Pi_{AA}(s)$,
is an order parameter of EWSB. In asymptotically-free gauge theories~\cite{Bernard:1975cd} it
vanishes as $1/s^3$, at large momenta. This implies UV super-convergence properties, the so-called Weinberg Sum Rules (WSRs) \cite{Weinberg:1967kj}:
\bel{eq:WSRs}
F_{V}^2 - F_{A}^2  = v^2\, ,\qquad\qquad
F_{V}^2  \,M_{V}^2 - F_{A}^2 \, M_{A}^2  = 0\, ,
\ee
which have been widely used in QCD \cite{Ecker:1988te,Ecker:1989yg}. In the electroweak case, they constrain the gauge boson self-energies~\cite{Peskin:1990zt}.
These two conditions determine $F_V$ and $F_A$ in terms of the resonance masses and imply $M_V < M_A$.
At the one-loop level, and together with \eqn{eq:FVGV} and \eqn{eq:FAkappa}, the WSRs imply also a relation between the Higgs gauge coupling and the resonance masses \cite{Pich:2012dv,Pich:2013fea}:
\bel{eq:kappa}
\kappa_W^{\phantom{2}}\, =\, M_V^2/M_A^2\, .
\ee
%

\begin{figure*}
\begin{minipage}[c]{5.5cm}
\includegraphics[width=5.5cm]{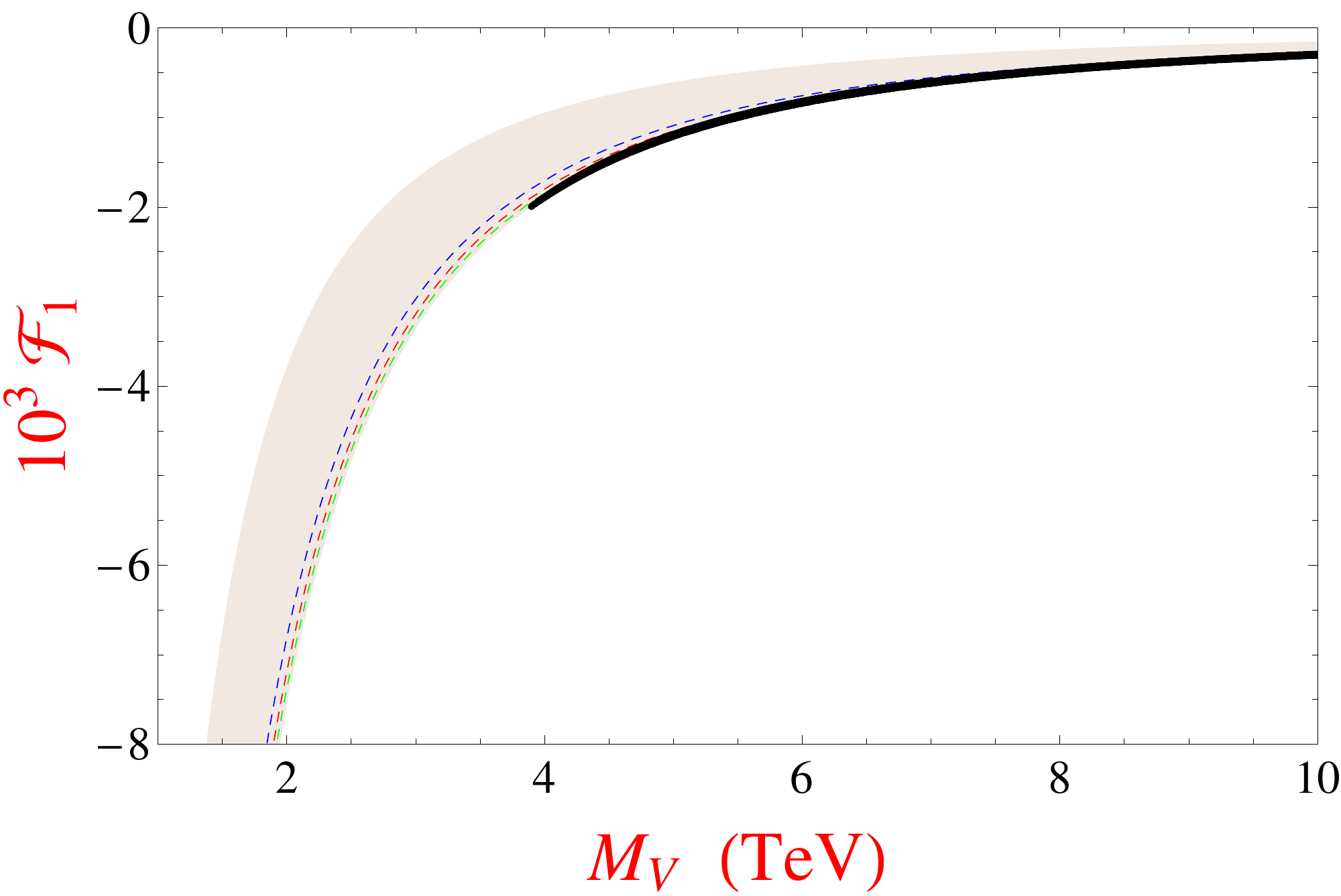}
\end{minipage}
\hskip .5cm
\begin{minipage}[c]{5.5cm}
\includegraphics[width=5.5cm]{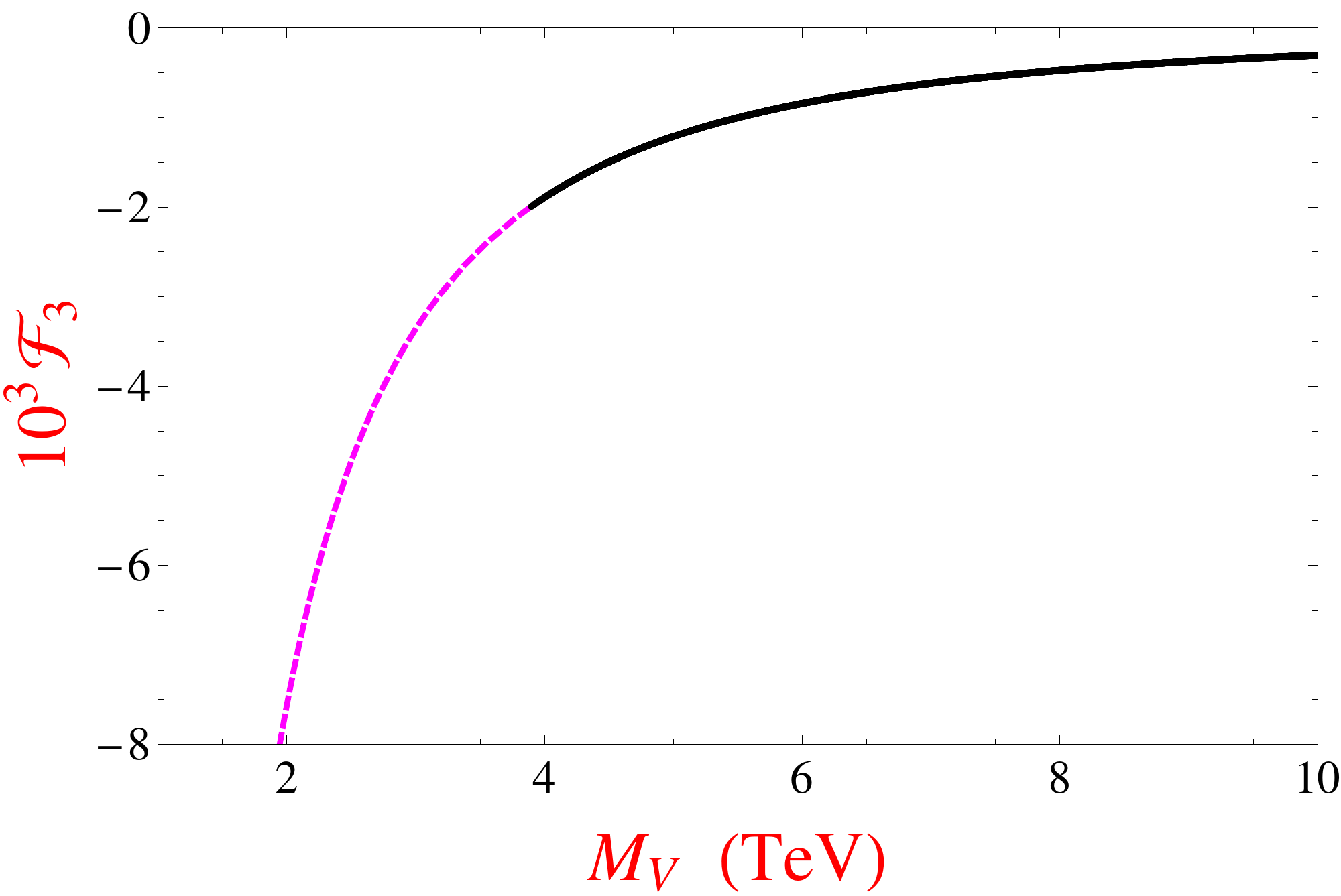}
\end{minipage}
\\[8pt]
\begin{minipage}[c]{5.5cm}
\includegraphics[width=5.5cm]{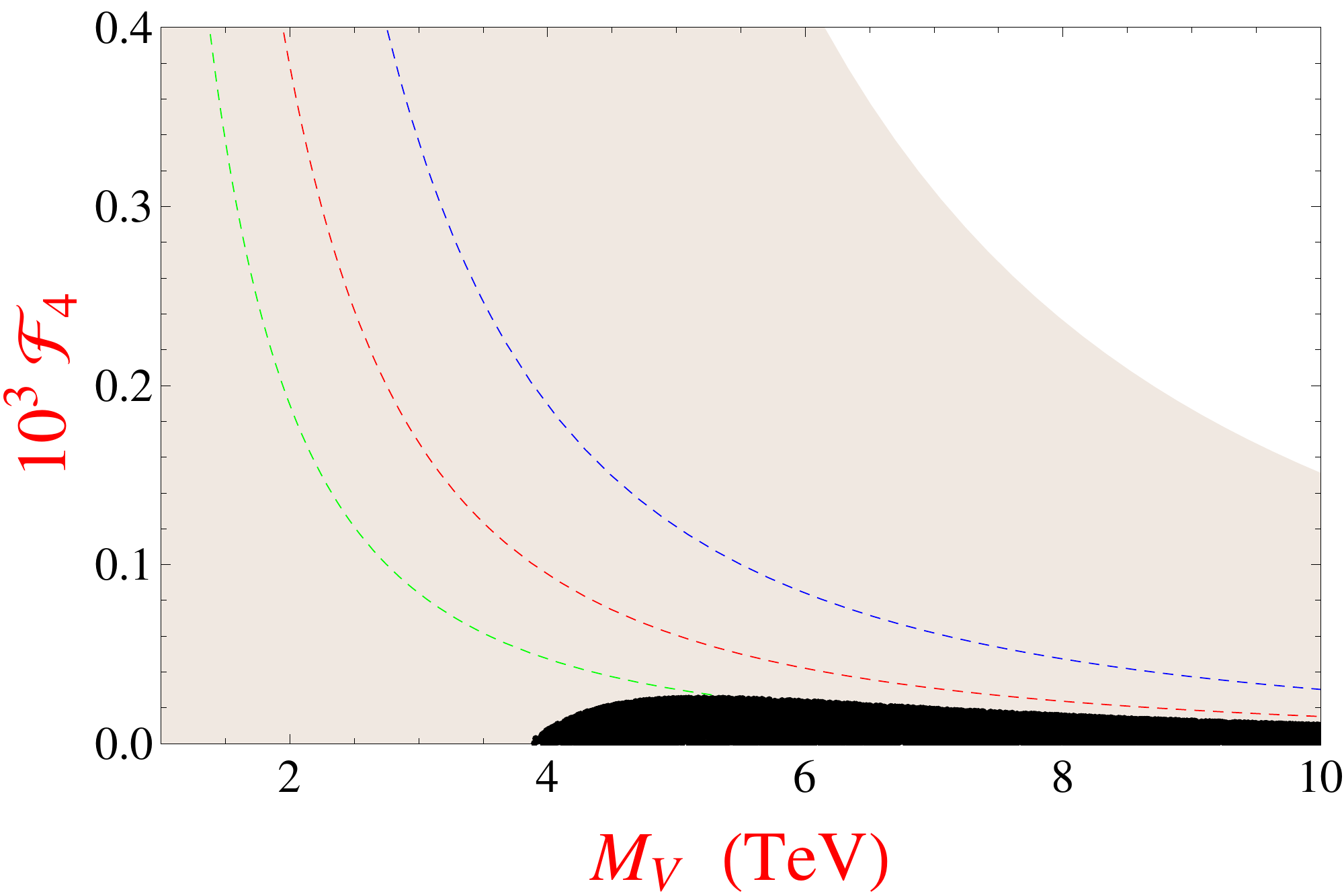}
\end{minipage}
\hskip .2cm
\begin{minipage}[c]{5.5cm}
\includegraphics[width=5.5cm]{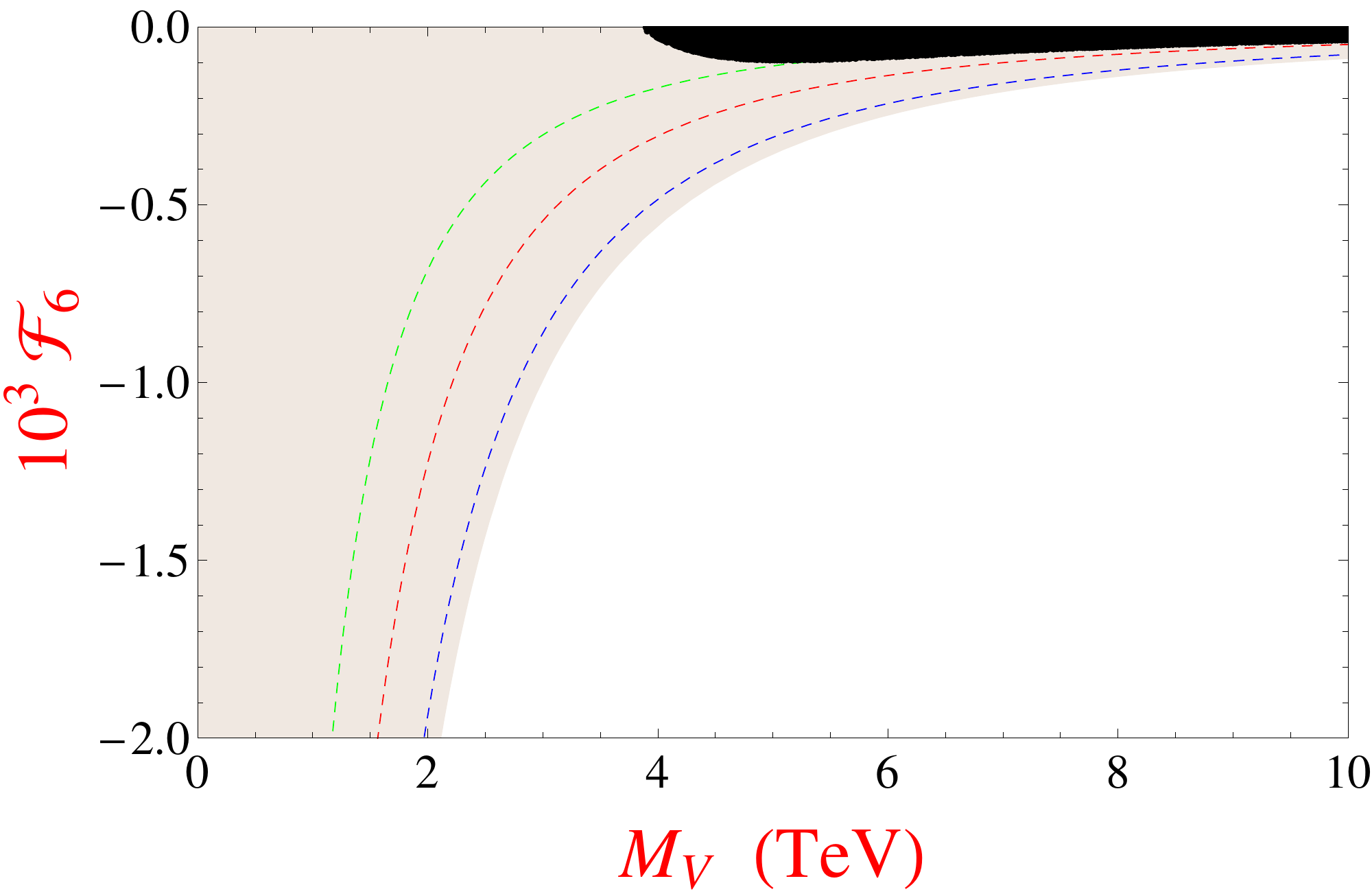}
\end{minipage}
\hskip .2cm
\begin{minipage}[c]{5.5cm}
\includegraphics[width=5.5cm]{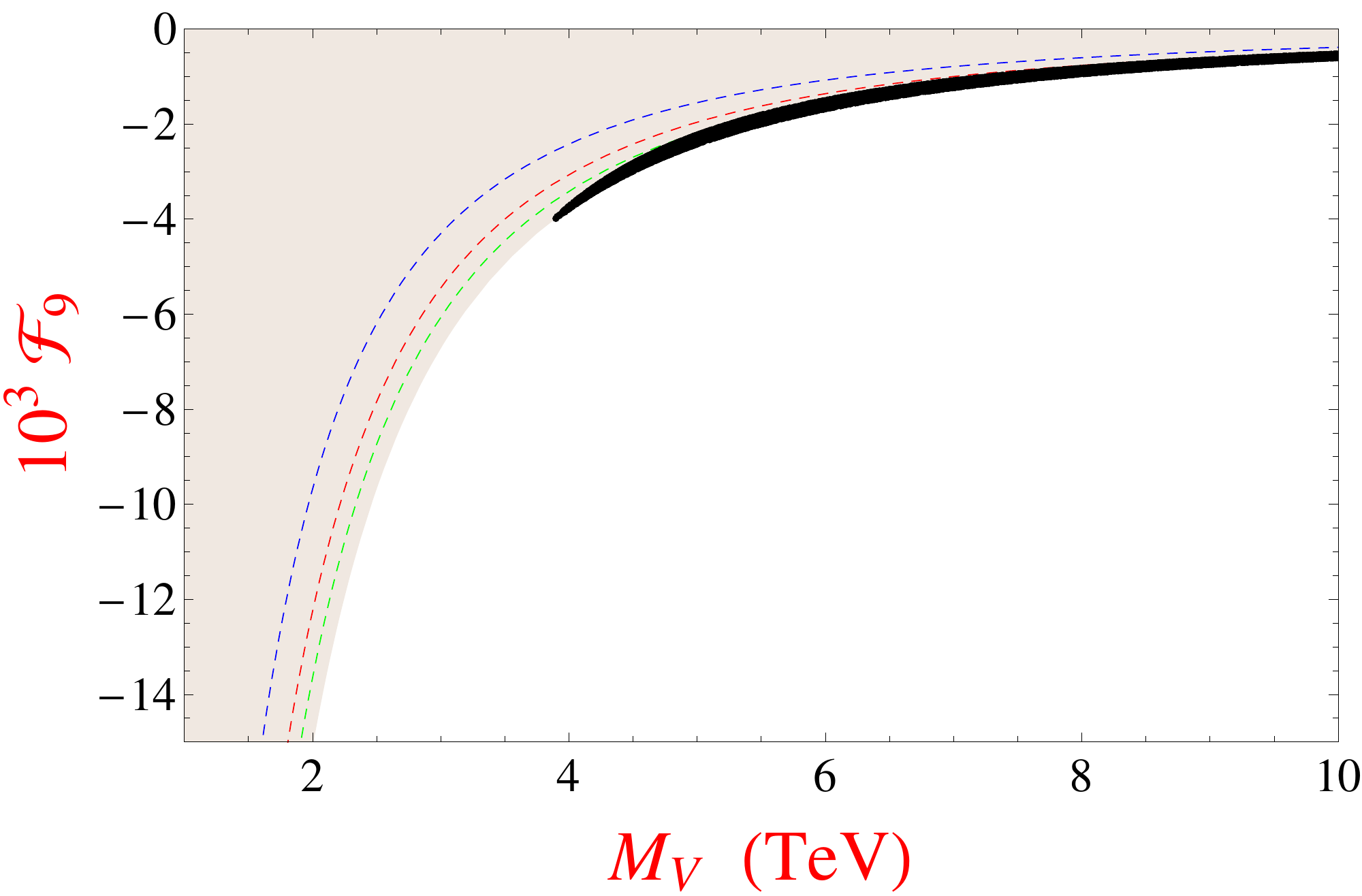}
\end{minipage}
\caption{Predicted $\mO(p^4)$ LECs for asymptotically-free theories,
as function of $M_V$. The light-shaded regions cover all possible values for $M_A>M_V$, while the blue, red and green lines correspond to $\kappa_W= M_V^2/M_A^2 = 0.8,\, 0.9$ and 0.95, respectively. $\mF_3$ does not depend on $M_A$. The oblique $S$ and $T$ constraints restrict the allowed ranges (95\% C.L.) to the dark areas.}
\label{fig:F1}
\vskip .4cm
\begin{minipage}[c]{5.5cm}
\includegraphics[width=5.5cm]{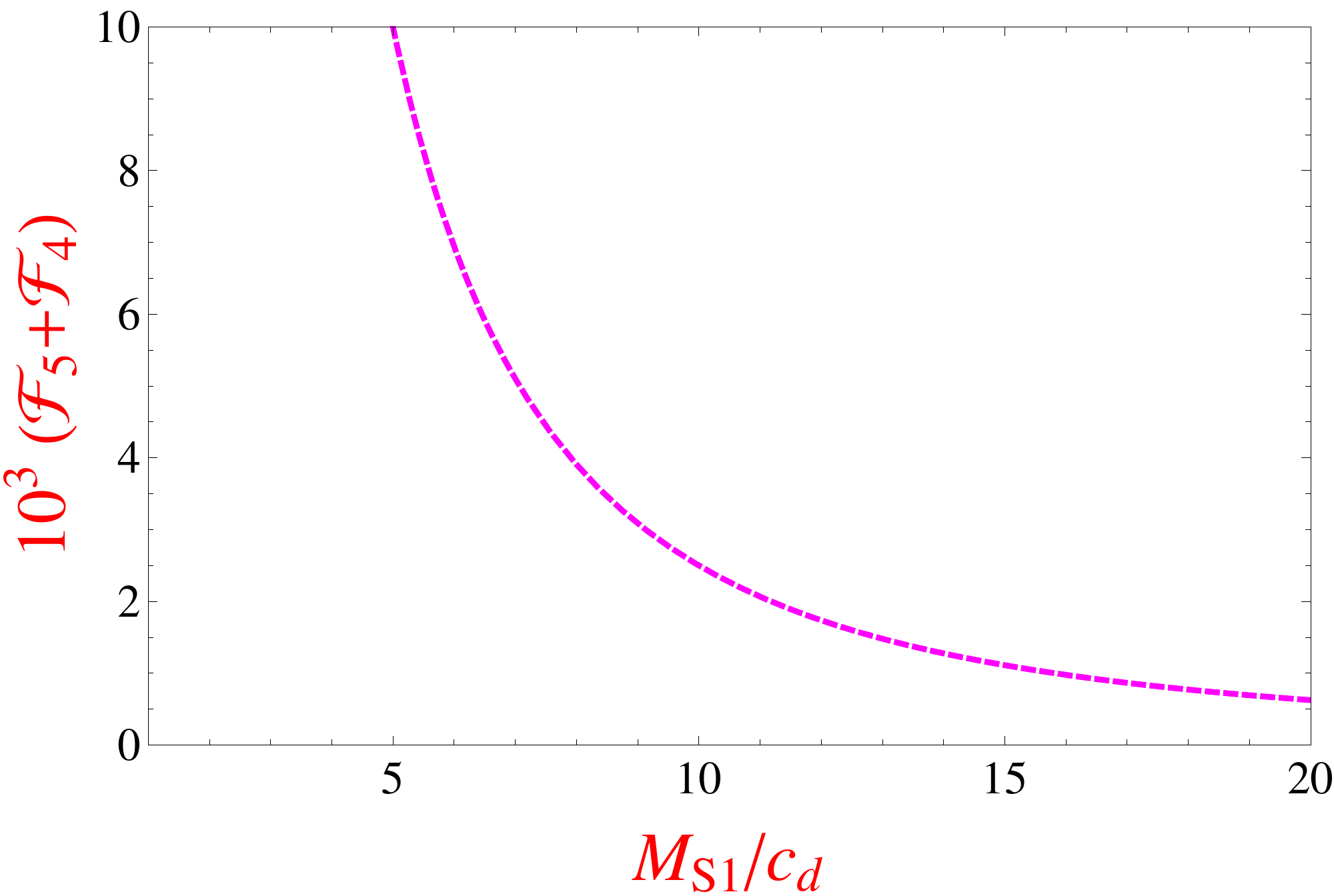}
\end{minipage}
\hskip .5cm
\begin{minipage}[c]{5.5cm}
\includegraphics[width=5.5cm]{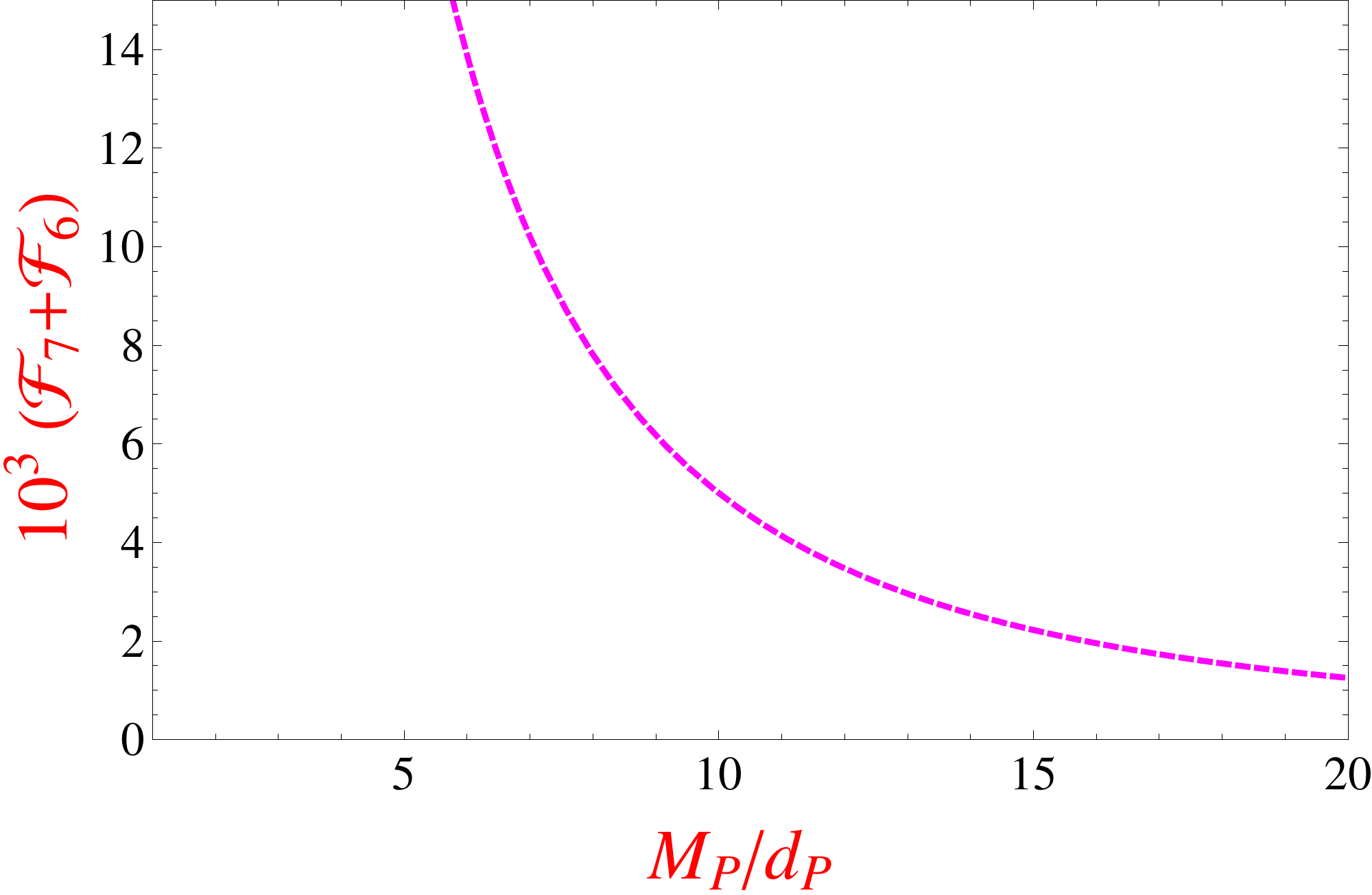}
\end{minipage}
\caption{Scalar and pseudoscalar contributions to $\mF_5$ and $\mF_7$, respectively.}
\label{fig:F2}
\end{figure*}

Using the identities \eqn{eq:FVGV}, \eqn{eq:FAkappa}, \eqn{eq:WSRs} and \eqn{eq:kappa}, all $\mO(p^4)$ LECs can be written in terms of $M_V$, $M_A$ and $v$, plus the scalar/pseudoscalar parameters entering $\mF_5$ and $\mF_7$. These improved predictions, shown in the r.h.s. expressions in Table~\ref{tab:LECs}, are valid in dynamical scenarios where the two WSRs are fulfilled, as happens in asymptotically-free theories. Softer conditions can be obtained imposing only the first WSR \cite{Pich:2012dv,Pich:2013fea,BigPaper}, which is also valid in gauge theories with non-trivial UV fixed points~\cite{Peskin:1990zt}.

Notice that the expressions derived in the previous section for the LECs are generic relations which include the functional dependence on $h/v$ hidden in the couplings. This is however no-longer true for our improved predictions incorporating short-distance constraints, where only constant couplings have been considered. Thus, the r.h.s. expressions in Table~\ref{tab:LECs} give the $\mO(h^0)$ term in the expansion of the corresponding LECs in powers of $h/v$.

In figure~\ref{fig:F1} the numerical values of the different LECs
$\mF_k\equiv \mF_k(0)$ are shown as functions of $M_V$, after imposing the short-distance constraints.
The light-shaded regions indicate all a priori possible values for $M_A > M_V$.
$\mF_2$ cannot be bounded with the current information and is not shown.
The dashed blue, red and green lines correspond to $\kappa_W=M_V^2/M_A^2 = 0.8,\, 0.9$ and 0.95, respectively. A single dashed purple curve is shown for $\mF_3$, which is independent of $M_A$.


The experimental bounds on the oblique parameters $S$ and $T$ imply that $\kappa_W > 0.94$ and $M_V> 4$~TeV (95\% C.L.), when the two WSRs apply~\cite{Pich:2012dv,Pich:2013fea}. The analysis of one-loop resonance contributions
to $S$ and $T$~\cite{Pich:2012dv,Pich:2013fea} restricts the allowed values of the LECs  to the dark areas shown in the figures, which results in quite strong limits (95\% C.L.):
$-2\times 10^{-3} <\mF_1<0$, $-2\times 10^{-3} <\mF_3<0$, $0<\mF_4< 2.5\times 10^{-5}$,
$-9\times 10^{-5}< \mF_6<0$, $-4\times 10^{-3}< \mF_9<0$.
These constraints would be softened in scenarios where only the 1st WSR applies~\cite{Pich:2012dv,Pich:2013fea}.

The exchanges of scalar and pseudoscalar resonances only manifest in $\mF_5$  and $\mF_7$.
These contributions can be isolated through the combinations $\mF_4+\mF_5$ and $\mF_6+\mF_7$ which depend only on the ratios $M_{S_1}/c_d$ and $M_P/d_P$, respectively.
The predicted values are shown in figure~\ref{fig:F2}.

\section{Summary}

We have analyzed in a model-independent way the low-energy implications of generic heavy resonances, from an underlying strongly-coupled dynamics, through their imprints on the LECs of the EWET. This is the only experimentally accessible information below the energy gap which separates the known particles from heavier new-physics states.

Integrating out the heavy fields, one gets the predictions given in Table~\ref{tab:LECs} in terms of resonance parameters. Adding mild assumptions about the UV dynamical behaviour,
valid in a broad variety of new-physics scenarios,
our results lead to very strong constraints \cite{BigPaper}, some of which are shown
in the r.h.s of Table~\ref{tab:LECs} and in figures~\ref{fig:F1} and \ref{fig:F2}.
The resulting pattern of LECs, with clear correlations characterizing the different quantum numbers of the massive states, will help to infer the type of short-distance physics underlying any deviation from the SM seen in future data.

The LECs $\mF_{1-5}$ induce anomalous gauge-boson couplings, $\mF_{1,3}$ are relevant for the oblique $S$ parameter \cite{Pich:2012dv,Pich:2013fea,Pich:2012jv} and the $\gamma\gamma\to WW$ amplitude~\cite{Delgado:2014jda,Chatrchyan:2013akv}, and
$\mF_{4-8}$ enter in $WW\to WW, hh$ scattering~\cite{Delgado:2013hxa,Espriu:2013fia,Aad:2014zda}.
At present, the experimentally most strongly constrained LEC is $\mF_1$ which gives a direct tree-level contribution to the oblique $S$ parameter, $\Delta S = -16\pi\,\mF_1$ (reflected in figure~\ref{fig:F1} at NLO). Future $e^+e^-$ colliders (LC/GigaZ) could bring a factor of 5 improvement in precision, reaching $\delta S\sim \pm 0.02$~\cite{Baak:2013fwa} or $\delta \mF_1\sim 4\cdot 10^{-4}$.

The combined analysis of LEP and collider data bounds the anomalous triple gauge-boson couplings in the range
$|\Delta g_1^Z|,\, |\Delta \kappa_Z|, \, |\Delta\kappa_\gamma| \lsim \cO(10^{-1})$~\cite{Falkowski:2015jaa,Corbett:2013pja}, which translates in a poor bound
$|\mF_3|\lsim \cO(10^{-1})$. While sizeable improvements are to be expected at the LHC, the LC could achieve~\cite{Baak:2013fwa,Baer:2013cma} $|\mF_3| \sim |\Delta\kappa_\gamma|/g^2\sim 5\cdot 10^{-4}$, becoming sensitive to the values predicted in figure~\ref{fig:F1}.

Regarding anomalous quartic gauge-boson couplings, LHC has provided the first evidences
on $WW$ scattering~\cite{Aad:2014zda},
giving bounds of the order of $|\mF_{4,5}|\lsim\cO(10^{-1})$.
Future LHC runs are expected to reduce the uncertainties on these two LECs
down to $\cO(10^{-3})$~ \cite{Fabbrichesi:2015hsa}. This is not enough to become
sensitive to vector-exchange contributions, but could allow us to pin down a scalar-singlet effect on $\mF_5$ in the range indicated in figure~\ref{fig:F2}.

The LECs $\mF_{6-9}$ involve the Higgs field in the external legs and are still unknown.
Some experimental information on these couplings will be obtained through single-Higgs and multi-Higgs production processes at the LHC and/or other future colliders.

\vspace{0.3cm}

\acknowledgments

{\bf Acknowledgments}: Work supported
by the Spanish Government and ERDF funds from the European Commission [FPA2011-23778, FPA2013-44773-P, FPA2014-53631-C2-1-P], by the Spanish Centro de Excelencia
Severo Ochoa Programme [SEV-2012-0249,   
SEV-2014-0398], the Generalitat Valenciana [PrometeoII/2013/007] and La Caixa [PhD grant for Spanish universities].


\end{document}